\documentclass[12pt]{article}

\makeatletter

\providecommand{\LyX}{L\kern-.1667em\lower.25em\hbox{Y}\kern-.125emX\@}

\usepackage{epsfig}
\usepackage{amsmath}
\usepackage{amssymb} 
\usepackage{cite}
\usepackage{latexsym} 
\usepackage{a4}
\usepackage{amsfonts}
\newcommand{\plabel}{\label}

\makeatother

\begin{document}

\begin{titlepage} \renewcommand{\thefootnote}{\fnsymbol{footnote}}

\hfill{}TUW--00--04 \\


{\par\centering \vspace{1cm}\par}

{\par\centering \textbf{\large Open strings, Born--Infeld action and the heat
kernel}\large \par}

{\par\centering \vspace{1.0cm} \vfill
\renewcommand{\baselinestretch}{1}\par}

{\par\centering \textbf{W.\ Kummer\( ^{1} \)\footnotemark[1] 
and D.V.\ Vassilevich\( ^{2} \)\footnotemark[2] \footnotemark[3]{}} \vspace{7ex}\par}

{\par\centering \( ^{1} \)Institut f\"{u}r Theoretische Physik, Technische
Universit\"{a}t Wien \\
 Wiedner Hauptstr. 8--10, A-1040 Wien \\
 Austria \vspace{2ex}\par}

{\par\centering \( ^{2} \)Institut f\"{u}r Theoretische Physik, Universit\"{a}t
Leipzig,\\
 Augustusplatz 10, D-04109 Leipzig,\\
 Germany\par}

{\par\centering \footnotetext[1]{e-mail: \texttt{wkummer@tph.tuwien.ac.at}} \footnotetext[2]{e-mail:
\texttt{Dmitri.Vassilevich@itp.uni-leipzig.de}} \footnotetext[3]{On leave from
Department of Theoretical Physics, St. Petersburg University, 198904 St. Petersburg,
Russia}\par}

\vfill

\begin{abstract}
In the derivation of the Born-Infeld action for the case with a 
nontrivial boundary of the string world sheet
the appearance of a new term changes the conformal anomaly. This may
have many consequences, especially also in the study of 
generalized interacting brane systems.
\end{abstract}
\end{titlepage}

\vfill

\section{Introduction}

An essential ingredient for the proper formulation of systems consisting of
strings and (D-) branes is the Born-Infeld (BI) action. The pioneering works
\cite{FT85,T86,ACNY87,CLNY88} derive it from the condition that
the beta-function for the string vanishes, when the string is embedded in
an external gauge field. From the extensive literature on the application of
this action in brane theory and its vast number of consequences \cite{T99}
we only refer to some recent work \cite{BOG96}which is also based upon older
results \cite{DLP,Nes}.

In our present note we address some delicate point in the derivation of the
BI action which is related to the existence of a nontrivial boundary. The use
of the heat kernel technique appears to be a necessity in this context. However,
in order to obtain a well-posed spectral problem it turns out that also a new
rule for Euclidean continuation has to be introduced. We also rely heavily on
the seminal paper of O. Alvarez \cite{Alvarez} who already a long time ago
strongly emphazised the consistent (gauge-invariant) treatment of the boundary.
In our present case these techniques are extended by the introduction of an
(abelian) gauge field.

\section{The sigma model}

To simplify the discussion we represent the string by a sigma model action with
Euclidean metric both on the world sheet and in space-time 
\begin{equation}
S=\frac{1}{2\pi \alpha' }\left[ \frac{1}{2}\int _{\cal M}d^{2}z\sqrt{h}
h^{ab}\partial _{a}X_{\mu }\partial _{b}X^{\mu }
+\int _{\partial {\cal M}}d\tau A_{\mu }\partial _{\tau }
X^{\mu }\right] \plabel {act}
\end{equation}
 The string tension \( 2\pi \alpha'  \) which is usually written as a factor
in front of the boundary term has been absorbed in \( A_{\mu } \) for convenience.
We suppose that the metric \( h_{ab} \) is flat, 
but the boundary \( \partial {\cal M} \)
can be of an arbitrary shape and may contain several connected components. We
ignore non-trivial background fields in the closed string sector 
(\( G_{\mu \nu }=\delta _{\mu \nu } \),)
and do not include a dilaton interaction and the \( B \)-field. 
The generalization
is straightforward for the case when these fields are taken into account. 
Let \( N^{a} \)
be an inward pointing unit normal to the boundary. We choose the coordinate
system in such way that on \( \partial {\cal M} \) the vectors 
\( N_{a}dz^{a},\, d\tau  \)
form an orthonormal pair.

It should be emphazised that the boundary term in (\ref{act}) is real in
contrast to the Euclidean actions of \cite{FT85,ACNY87}. There the relative
factor of \( i \) originated from the continuation of the volume element \( \sqrt{-h}\rightarrow i\sqrt{h} \)
in the first term. We suggest to also rotate \( A_{\mu } \) to \( iA_{\mu } \)
during continuation to the Euclidean space so that the factor \( i \) is compensated.
Actually, it does not matter which particular continuation is chosen as long
as one gets a meaningful theory in Euclidean space so that a proper continuation
back to the physical Minkowski space is possible. As will be 
demonstrated below,
our rule of continuation has the crucial advantage to provide a well-posed spectral
problem in Euclidean space, as opposed to the usual approach \cite{FT85,ACNY87}
when the factor \( i \) enters the boundary conditions for real fields and
thus makes the boundary value problem ill-defined. The fact that parity odd
fields can get an imaginary factor in the Euclidean space has been observed
long ago \cite{AB} in the context of chiral theories. It is easy to see that
the field \( A_{\mu } \) is of odd parity from the world volume point of view.
Indeed, the last term in (\ref{act}) can be rewritten as 
\( \int _{\cal M}\partial _{a}
(\varepsilon ^{ab}A_{\mu }\partial _{b}X^{\mu }) \).
Here \( A_{\mu } \) couples to the parity-odd quantity \( \varepsilon ^{ab} \).
Whenever it is possible to compare our results to those of previous related
papers \cite{FT85,ACNY87} where this mathematical subtlety was just ignored,
they are compatible after the replacement \( A\rightarrow iA \).

Let us expand the action (\ref{act}) around an arbitrary background \( \bar{X} \),
\( X=\bar{X}+\xi  \). To calculate the one-loop (in the field theory sense)
effective action we need only the part which is quadratic in \( \xi  \): 
\begin{eqnarray}
 &  & S_{2}=\frac{1}{2\pi \alpha' }\left[ \frac{1}{2}
\int _{\cal M}d^{2}z\sqrt{h}h^{ab}\partial _{a}\xi _{\mu }\partial _{b}
\xi ^{\mu }+\right. \nonumber \\
 &  & \qquad +\left. \frac{1}{2}
\int _{\partial {\cal M}}\left( F_{\nu \mu }
\xi ^{\nu }\dot{\xi }^{\mu }+\dot{\bar{X}}^{\mu }
\partial _{\nu }F_{\rho \mu }\xi ^{\nu }\xi^{\rho }\right) 
\right] \label {S2}
\end{eqnarray}
 Clearly the one-loop effective action is 
\begin{equation}
W=\frac{1}{2}\log \det (-\Delta \delta _{\mu \nu })\,,\plabel {W1}
\end{equation}
 with the scalar Laplacian \( \Delta  \) and with the boundary condition 
\begin{equation}
(-\partial _{N}\delta _{\mu \nu }+F_{\mu \nu }
\partial _{\tau }+\dot{\bar{X}}^{\rho }
(\partial _{\nu }F_{\mu \rho }))\xi ^{\nu }|_{\partial {\cal M}}=0\, 
\, ,\plabel {bc1}
\end{equation}
 where \( \partial _{N}=N^{a}\partial _{a} \) is the derivative with respect
to the inward pointing unit normal vector N. 
It is useful to rewrite (\ref{bc1})
by adding and subtracting a term with \( \mu \leftrightarrow \nu  \) as 
\begin{equation}
{\cal B}\xi |_{\partial {\cal M}}=
\left( \partial _{N}+\frac{1}{2}(\partial _{\tau }\Gamma 
+\Gamma \partial _{\tau })+S\right) \xi |_{\partial {\cal M}}=0
\,, \label {bc2}
\end{equation}
 where 
\begin{equation}
\Gamma _{\mu \nu }=-F_{\mu \nu },\qquad S_{\mu \nu }=-\frac{1}{2}\dot{\bar{X}}^{\rho }(\partial _{\nu }F_{\mu \rho }+\partial _{\mu }F_{\nu \rho })\, . 
\plabel {SGam}
\end{equation}
 Thus the boundary condition (\ref{bc2}) also ensures the hermiticity of the
Laplace operator. Indeed, 
\begin{equation}
\label{herm}
\int _{\cal M}\left( {\xi' }_{\mu }\Delta \xi ^{\mu }
-{\xi }_{\mu }\Delta {\xi' }^{\mu }\right) 
=\int _{\partial {\cal M}}\left( -{\xi' }_{\mu }\partial _{N}\xi ^{\mu }
+{\xi }_{\mu }\partial _{N}{\xi' }^{\mu }\right) =0\, ,
\end{equation}
 if \( \xi  \) and \( \xi'  \) satisfy (\ref{bc2}), when the hermiticity
of \( \frac{1}{2}(\partial _{\tau }\Gamma +\Gamma \partial _{\tau })+S  \) 
is used. The boundary condition (\ref{bc2}) contains
tangential derivatives and belongs to the Gilkey-Smith class\footnote{%
Sometimes the boundary conditions with tangential derivatives are called ``mixed''
\cite{ASJ}. In the theory of the heat equation asymptotics the name ``mixed''
is reserved for a completely different type of the boundary conditions \cite{Gilkey}.
}.

\section{The heat kernel expansion}

In this section we collect some basic information on the heat kernel expansion
for Gilkey--Smith boundary conditions. The spectral geometry of such boundary
conditions was first studied in \cite{GilkeySmith,McO}. The increased interest
for that heat kernel expansion with these boundary conditions in recent years
was motivated primarily by one-loop calculations in quantum cosmology 
\cite{AEearlier,DK,AE,EV,AvEs}.

Consider an operator of the Laplace type \( D=-(\nabla ^{a}\nabla _{a}+E) \)
acting in a smooth vector bundle over a smooth Riemannian manifold 
\( \mathcal{M} \) of
dimension \( m \). \( \nabla  \) denotes a covariant derivative with respect
to a certain connection on \( \mathcal{M} \). \( E \) is an endomorphism (a
matrix-valued function). Let the boundary operator be 
\begin{equation}
{\mathcal{B}}=\nabla _{N}+\frac{1}{2}
(\Gamma ^{i}\hat{\nabla }_{i}
+\hat{\nabla }_{i}\Gamma ^{i})+S\, \, ,\plabel {boundop}
\end{equation}
 where \( \Gamma ^{i} \) and \( S \) are some matrices 
depending on the coordinates
on the boundary, \( i=1,\dots ,m-1 \). The covariant 
derivative \( \hat{\nabla } \)
contains both the standard Levi-Civita connection of 
the boundary and the restriction
of the bundle connection to the boundary. The operator (\ref{boundop}) defines
the boundary condition \( {\mathcal{B}}\phi |_{\partial M}=0 \). The Laplace
operator \( D \) is symmetric if the \( \Gamma ^{i} \)'s are antihermitian
and \( S \) is hermitian. In our case the matrices (\ref{SGam}) satisfy this
requirement.

If \( f \) is a smooth function on \( \mathcal{M} \), there is an asymptotic
series as \( t\rightarrow +0 \) of the form 
\begin{equation}
{\textrm{Tr}}(f\exp (-tD))=
\sum _{n\geq 0}t^{-m/2+n}a_{n}(f,D,{\mathcal{B}})\,,\plabel {asymp}
\end{equation}
 where \( n=0,\frac{1}{2},1,\dots  \).  The functional
trace in the space of square integrable functions is denoted by ${\rm Tr}$. 
The heat
kernel (\ref{asymp}) is well defined only if the boundary value problem is
strongly elliptic. A criterion for strong ellipticity for the boundary 
conditions
(\ref{boundop}) has been proved recently in \cite{AvEs}. Roughly speaking,
it requires the absolute values of the eigenvalues of \( \Gamma ^{i} \) to
be smaller than \( 1 \). In the present context the points where the strong
ellipticity is lost correspond to critical values of the electric field, which
were discussed for Minkowski signature in \cite{Nes}.

The bulk terms in the heat kernel expansion do not depend on \( \Gamma  \).
The leading boundary contributions to the heat kernel can be taken directly
from refs. \cite{McO,AE}
\begin{eqnarray}
 &  & a_{1/2}(f,P,{\mathcal{B}})=\frac{1}{(4\pi )^{(m-1)/2}}
\int _{\partial {\cal M}}{\textrm{tr}}(\gamma f)\,,\nonumber \\
 &  & a_{1}(f,P,{\mathcal{B}})=\frac{1}{(4\pi )^{m/2}}
\int _{\partial {\cal M}}{\textrm{tr}}\left( f(b_{0}k+b_{2}S
+\sigma _{1}k_{ij}\Gamma ^{i}\Gamma ^{j})+b_{1}\nabla _{N}f\right) \,,
\plabel {a1}
\end{eqnarray}
 where \( k_{ij} \) is the extrinsic curvature of the boundary. 
Here \( {\textrm{tr}} \)
is the ordinary matrix trace. The functions 
\( b_{0},\, b_{1},\, b_{2},\, \gamma ,\, \sigma _{1} \)
are 
\begin{eqnarray}
 &  & \gamma =\frac{1}{4}\left[ \frac{2}{\sqrt{1+\Gamma ^{2}}}-1\right]\,, 
\nonumber \\
 &  & b_{1}=\frac{1}{\sqrt{-\Gamma ^{2}}}{\textrm{Artanh}}
(\sqrt{-\Gamma ^{2}})-\frac{1}{2}\,,\nonumber \\
 &  & b_{2}=\frac{2}{1+\Gamma ^{2}}\,,\nonumber \\
 &  & b_{0}+\sigma _{1}\Gamma ^{2}=\frac{1}{3}\, \,, \label {funcs}
\end{eqnarray}
 In two dimensions the functions \( b_{0} \) and \( \sigma _{1} \) enter the
heat kernel expansion only in the combination 
\( b_{0}+\sigma _{1}\Gamma ^{2} \)
due the identity \( k_{ij}\Gamma ^{i}\Gamma ^{j}=k\Gamma ^{2} \), valid on the
one-dimensional boundary.

In this paper we are not discussing the contribution of the zero modes to the
path integral. Therefore, strictly speaking the 
equations (\ref{a1}) and (\ref{funcs})
are valid only up to some ``global'' effects.

\section{Beta function and the conformal anomaly}

Next we make use of the \( \zeta  \)-function regularization \cite{zeta}.
The zeta function of an elliptic operator \( D \) is defined as 
\begin{equation}
\zeta _{D}(s)={\textrm{Tr}}(D^{-s})\, \, .\plabel {defzeta}
\end{equation}
 In term of the zeta function (\ref{defzeta}) the effective action (\ref{W1})
reads 
\begin{equation}
W=-\frac{1}{2s}\zeta _{D}(0)-\frac{1}{2}\zeta' _{D}(0)\,,\label {W2}
\end{equation}
 where the prime denotes differentiation with respect to \( s \).

By using the well known relation between the zeta function and the heat kernel
coefficients \( \zeta _{D}(0)=a_{1}(1,D,{\mathcal{B}}) \) the divergent part
of the effective action at \( s\rightarrow 0 \) may be written as 
\begin{equation}
W_{{\mbox {\scriptsize {div}}}}=-\frac{1}{2s}\frac{1}{4\pi }
\int _{\partial {\cal M}}d\tau \, 
\left[ -\dot{\bar{X}}^{\rho }(\partial _{\nu }F_{\mu \rho }
+\partial _{\mu }F_{\nu \rho })(1+F^{2})^{-1}_{\nu \mu }
+\frac{1}{3}k\delta _{\nu }^{\nu }\right] .\label {Wdiv}
\end{equation}
 The first term on the right hand side of (\ref{Wdiv}) can be represented as
\( (1/2\pi )\int _{\partial {\cal M}}d\tau \, G_{\mu }\dot{\bar{X}}^{\mu } \). From
this we can read off the beta function (in the notations of \cite{ACNY87}) 
\begin{equation}
\beta _{\mu }^{A}\propto (\partial _{\rho }F_{\nu \mu })(1+F^{2})^{-1}_{\nu \rho }\,.\label {beta}
\end{equation}
 After collecting the factors of \( i \) which originate
from our prescription for Euclidean continuation
 it gives the same equation of motion as in \cite{ACNY87} and,
therefore, reproduces the variation of the BI action \cite{FT85} 
for the \( A_{\mu } \).
It should stressed that to derive this result we neither supposed 
a special geometry
of the world sheet, nor had to
neglect higher derivatives of \( F_{\mu \nu } \).

The second term under the integral in (\ref{Wdiv}), which is proportional to
the dimension of the target space \( \delta _{\nu }^{\nu } \), does not depend
on \( F \) and is always present in the theory of open strings \cite{Alvarez}.
Since we had assumed that the scalar curvature of the 
two-manifold \( \mathcal{M} \)
is zero, this term can be expressed in terms of the Euler characteristic of
\( {\cal M} \): \( 2\pi \chi (\mathcal{M})=\int _{\partial {\cal M}}d\tau k \).

We next turn to the conformal anomaly. An infinitesimal conformal 
transformation
\( \delta h_{ab}=(\delta k)h_{ab} \) with a local parameter \( \delta k \)
produces the trace of the (effective) energy momentum tensor 
\begin{equation}
\delta W_{{\mbox {\scriptsize {ren}}}}=
\frac{1}{2}\int _{\cal M}d^{2}z\sqrt{h}\delta h^{ab}T_{ab}
=-\frac{1}{2}\int _{\cal M}d^{2}z\sqrt{h}\delta k(x)T_{a}^{a}(x)\,,\label {T}
\end{equation}
 where the \( W_{{\mbox {\scriptsize {ren}}}} \) is the second (finite) term
in (\ref{W2}). It is quite important that both the Laplace operator and the
boundary operator transform covariantly under the metric rescaling 
\begin{eqnarray}
 &  & \Delta \rightarrow (1-k+\dots )\Delta ,\label {delD}\\
 &  & {\mathcal{B}}\rightarrow (1-\frac{k}{2}+\dots ){\mathcal{B}}\,.
\label {delB}
\end{eqnarray}
 The second property (\ref{delB}) follows from our assumption that 
\( N_{a}dz^{a},d\tau  \)
are two orthonormal vectors -- which we used to derive the equation 
(\ref{bc1}).
It guarantees that the functional space defined by (\ref{bc1}) is invariant
under the conformal transformations.

With the definition of a generalized \( \zeta  \)-function 
\begin{equation}
\zeta (s|\delta k,D)={\textrm{Tr}}(\delta kD^{-s})\plabel {varW}
\end{equation}
 the variation in (\ref{T}) can be identified with 
\begin{equation}
\plabel {TX}\delta W_{{\mbox {\scriptsize {ren}}}}=-\frac{1}{2}\zeta (0|\delta k,D)\quad ,
\end{equation}
 where we used \( \delta \zeta _{D_{k}}(s)=s\mbox {Tr}(D^{-s}\delta k) \).
Combining (\ref{TX}) and (\ref{T}) we obtain 
\begin{equation}
\zeta (0|\delta k,D)=\int d^{2}z\sqrt{h}\delta k(z)T_{a}^{a}(x)\; .\plabel {T2}
\end{equation}
 By a Mellin transformation one can show that \( \zeta (0|\delta k,D)=a_{1}(\delta k,D,{\mathcal{B}}) \).
The (smeared) conformal anomaly reads: 
\begin{eqnarray}
 &  & \int _{\cal M}\sqrt{h}d^{2}zf(z)T_{a}^{a}(z)=
\frac{1}{4\pi }\int _{\partial {\cal M}}d\tau \, 
\left[ f(\tau )\left( \frac{1}{3}k\delta _{\nu }^{\nu }
-2\dot{\bar{X}}^{\rho }(\partial _{\nu }F_{\mu \rho })
(1+F^{2})^{-1}_{\nu \mu }\right) \right. \nonumber \\
 &  & \qquad \qquad \left. +(\nabla _{N}f)\left( 
(-F^{2})_{\mu \nu }^{-1/2}{\textrm{Artanh}}(\sqrt{-F^{2}})_{\nu \mu }-
\frac{1}{2}\delta _{\mu }^{\mu }\right) \right]\,. \label {anomaly}
\end{eqnarray}
 In the limit \( F\rightarrow 0 \) the conformal anomaly (\ref{anomaly}) 
coincides
with the standard expression \cite{Alvarez}. The second term on the first line
appears quite naturally and is a manifestation of the BI action. The second
line contains a somewhat unusual contribution. To the best of our knowledge
it has not been obtained before. To perform a full-scale analysis of this term
one should include the dilaton field in the ``bare'' action (\ref{act}).
We postpone this to a future more detailed publication. In any case, the last
term in (\ref{anomaly}) suggests a very interesting interplay between the gauge
sector and the conformal sector of string theory.

\section{Conclusions}

By a careful interpretation of the transition to Euclidean space we are able
to reformulate the problem of the string in the presence of a nontrivial 
boundary
where the string is coupled to an abelian background field. As a consequence
of the effect that our operator, appearing in the boundary condition, 
is hermitian
we arrive at a well-posed elliptic problem in the sense 
of the heat kernel technique
where known formulas from that field can be applied directly. 
Our central result
is the one for the conformal anomaly at the boundary, Eq. (\ref{anomaly}). 
Beside
the standard term leading to the BI action for the gauge field from the 
vanishing
of the beta-function and a term proportional to the extrinsic curvature at the
boundary (which was known for a long time \cite{Alvarez}) we find a 
new contribution
which depends on the gauge field \( F_{\mu \nu } \). It should be stressed
that our whole argument (in contrast to previous ones for the BI action) 
contains
no restriction on the vanishing of derivatives for \( F_{\mu \nu } \) . Also
no special geometry of the world volume need be assumed. This means that
our calculations are valid for an arbitrary number of string loops.
Of course, higher loop corrections (in the sense of quantum
field theory) will contain higher derivatives of $F_{\mu\nu}$
\cite{AT88}. 

We see a rather wide range of applications of our present result, the 
most obvious extension being the presence of other fields \cite{CFMP85,T98}
which, however,
should not present any new
technical difficulties. Also the indication for a further contribution to the
anomaly at the intersection of branes possibly could shed new light upon the
problems encountered for interacting strings and branes 
\cite{brane}.
There could be even a relation to the very recent work on the noncommutative
geometry approach for strings and branes \cite{SW99}.

\section*{Acknowledgments}

We are grateful to A. Andrianov, I. Avramidi, I. Bandos and G. Esposito for
discussions and/or correspondence. This work has been supported by the Fonds
zur F\"{o}rderung der wissenschaftlichen Forschung project P-12.815-TPH. One
of the authors (D.V.) thanks the Alexander von Humboldt Foundation and 
the Russian
Foundation for Fundamental Research, grant 97-01-01186, for the support.

\section*{Note added}

After this paper has been completed and posted on the net
we were informed
by Professor Osborn that our expression for the
conformal anomaly (22) (that is our main result)
is contained in the Appendix
of his paper \cite{Osborn}. We find it however quite
striking that such important result is not widely
recognised in modern literature on strings and
branes. Therefore, we decided to leave our preprint
on the net and add this short note.


\begin{thebibliography}{10}
\bibitem{FT85}E.S. Fradkin and A.A. Tseytlin, 
Phys. Lett. \textbf{163 B} (1985) 123. 
\bibitem{T86} A.A. Tseytlin, Nucl. Phys. \textbf{B 276} (1986) 391. 
\bibitem{ACNY87} A. Abouelsaood, C.G. Callan, C.R. Nappi and S.A. Yost, 
Nucl. Phys. \textbf{B
280 {[}FS 18{]}} (1987) 599. 
\bibitem{CLNY88} C.G. Callan, C. Lovelace, C.R. Nappi and S.A. Yost, 
Nucl. Phys. \textbf{B 308}
(1988) 221. 
\bibitem{T99}A.A. Tseytlin, Born-Infeld action, 
supesymmtery and string theory, hep-th/9908105,
to appear in the Yuri Golfand memorial volume, ed. M. Shifman, 
(World Scientific,
2000).
 
\bibitem{BOG96}A.A. Bytsenko, S.D. Odintsov and L.N. Granda, 
Phys. Lett. \textbf{B 387} (1996)
282. 
\bibitem{DLP}J. Dai, R.G. Leigh and J. Polchinski, 
Mod. Phys. Lett. \textbf{A 4} (1989) 2073;
R. Leigh, Mod. Phys. Lett. \textbf{A 4} (1989) 2767. 
\bibitem{Nes}
V.V. Nesterenko, Int. J. Mod. Phys. \textbf{A 10} (1989) 2627.
\bibitem{Alvarez}O. Alvarez, Nucl. Phys. \textbf{B 216} (1983) 125. 
\bibitem{AB} A.A. Andrianov, L. Bonora and R. Gamboa-Saravi, 
Phys. Rev. \textbf{D 26} (1982)
2821; A.A. Andrianov and L. Bonora, 
Nucl. Phys. \textbf{B 233} (1984) 232; 247. 
\bibitem{ASJ} H. Arfaei and M.M. Sheikh Jabbari, 
Nucl. Phys. \textbf{B 526} (1998) 278. 
\bibitem{Gilkey} P.B. Gilkey, 
\textit{Invariance Theory, the Heat Equation, and the Atiyah-Singer
Index Theorem}, (CRC Press, Boca Raton, 1994). 
\bibitem{GilkeySmith} P.B. Gilkey and L. Smith, J. Diff. Geom. 
\textbf{15} (1983) 393. 
\bibitem{McO} D.M. McAvity and H. Osborn, Class. Quantum Grav. 
\textbf{8} (1991) 1445. 
\bibitem{DK} J.S. Dowker and K. Kirsten, Class. Quantum Grav. 
\textbf{14} (1997) L169; \textbf{16}
(1999) 1917.
\bibitem{AEearlier} I. Avramidi and G. Esposito, Class. Quantum Grav. 
\textbf{15} (1998) 1141. 
 
\bibitem{AE} I. Avramidi and G. Esposito, Class. Quantum Grav. 
\textbf{15} (1998) 281. 
\bibitem{EV} E. Elizalde and D.V. Vassilevich, 
Class. Quantum Grav. \textbf{16} (1999) 813. 
\bibitem{AvEs} I.G. Avramidi and G. Esposito, 
Commun. Math. Phys. \textbf{200} (1999) 495;
Trends in Mathematical Physics: Proceedings. 
Edited by V. Alexiades and G. Siopsis.
Cambridge, Mass., International Press, 1999. 
(AMS/IP Studies in Advanced Mathematics,
v. 13). pp. 15-34. 
 
\bibitem{zeta}J.S. Dowker and R. Critchley, Phys. Rev. 
\textbf{D 13} (1976) 3224; S.W. Hawking,
Commun. Math. Phys. \textbf{55} (1977) 133. 
\bibitem{AT88}
O.D. Andreev and A.A. Tseytlin, Nucl. Phys. {\bf B 311} (1988) 205;
Mod. Phys. Lett. {\bf A 3} (1988) 1349.

\bibitem{CFMP85}C.G. Callan, D. Friedan, E.J. Martinec and M.J. Perry, 
Nucl. Phys. \textbf{B 262} (1985) 593. 
\bibitem{T98}A.A. Tseytlin, Nucl. Phys. \textbf{B 524} (1998) 41.

\bibitem{brane}
G. Papadopoulos, P.K. Townsend, Phys. Lett. \textbf{B 380} (1996) 273;
E. Bergshoeff, M. de Roo, E. Eyras, B. Janssen, J.P. van der Schaar, 
Nucl. Phys.
\textbf{B 494} (1997) 119;  P.M. Cowdall, P.K. Townsend, 
Phys. Lett. \textbf{B 429}
(1998) 281; \textbf{B 434} (1998) 458 (E); M. Cederwall, 
Mod. Phys. Lett. \textbf{A
12} (1997) 2641; Ph. Brax, J. Mourad, Phys. Lett \textbf{B 408}
(1997) 142; \textbf{B 416} (1998) 295; T. Sato, Phys. Lett. \textbf{B 439} 
(1998)
12; \textbf{B 441} (1998) 105; E. Bergshoeff, R. Kallosh, T.
Ortin, G. Papadopoulos, Nucl. Phys. \textbf{B 502} (1997) 149; J. Gomis, D.
Mateos, J. Simn, P.K. Townsend, Phys. Lett. \textbf{B 430} (1998) 231; 
C.S. Chu, E. Sezgin, JHEP \textbf{9712} (1997) 001; C.S. Chu,
P.S. Howe, E. Sezgin, Phys. Lett \textbf{B 428} (1998) 59; C.S.
Chu, P.S. Howe, E. Sezgin, P.C. West, Phys. Lett. \textbf{B 429} (1998) 273;
J.P. Gauntlett, N.D. Lambert, P.C. West, Commun. Math. Phys.
\textbf{202} (1999) 571;  
E. Sezgin, Topics in M-Theory, in Proceedings of the Abdus
Salam Memorial Meeting, ed. J. Ellis et al (World Scientific,
Singapore, 1999) hep-th/9809204;
P. West, Supergravity, Brane Dynamics and String Duality, hep-th/9811101;
I. Bandos and W. Kummer, Phys. Lett. {\bf B 462} (1999) 254;
Superstrings ``ending'' on super D-9 brane, hep-th/9906041.
\bibitem{SW99} N. Seiberg and E. Witten, 
JHEP \textbf{9909} (1999) 032. 
\bibitem{Osborn}
H. Osborn, Nucl. Phys. {\bf B 363} (1991) 486.


\end{thebibliography}
\end{document}